# Survey on Individual Differences in Visualization

Zhengliang Liu[1], R. Jordan Crouser[2], and Alvitta Ottley[1]

[1]Washington University in St. Louis, USA
[2]Smith College, USA

## Abstract

*Developments in data visualization research have enabled visualization systems to achieve great general usability and application across a variety of domains. These advancements have improved not only people's understanding of data, but also the general understanding of people themselves, and how they interact with visualization systems. In particular, researchers have gradually come to recognize the deficiency of having one-size-fits-all visualization interfaces, as well as the significance of individual differences in the use of data visualization systems. Unfortunately, the absence of comprehensive surveys of the existing literature impedes the development of this research. In this paper, we review the research perspectives, as well as the personality traits and cognitive abilities, visualizations, tasks, and measures investigated in the existing literature. We aim to provide a detailed summary of existing scholarship, produce evidence-based reviews, and spur future inquiry.*

## 1. Introduction

The term *individual differences* refers to individuals' "traits or stable tendencies to respond to certain classes of stimuli or situations in predictable ways" [DW96]. Much of the literature on individual differences has roots in psychology. Psychological research has demonstrated that people with distinct personality types and various cognitive abilities exhibit observable differences in task-solving and behavioral patterns [WB00, Ajz05]. Studies dating back to the late 1920s began by investigating variations in workplace performance [Hul28]. Throughout the intervening century, these findings have been extended to investigate individual characteristics that may predict performance under various conditions.

In the past few decades, the computational sciences have begun to recognize the role individual differences might play in shaping interaction in human-machine systems. For example, Benyon and Murray observed a relationship between *spatial ability* (a metric that measures a personâĂŹs ability to mentally represent and manipulate two- or three-dimensional objects) and task performance and preferences when using common interaction paradigms such as menus and the command line [BM93]. Nov et al. [NALB13] found that *extraversion* (one's tendency to engage with the external world) and *neuroticism* (a measure of emotional stability) had effects on users' contributions to online discussions, and suggested adaptations to certain visual cues to cater to different personality types. Gajos and Chauncey [GC17] observed that *introverted* people were more likely to use adaptive features in user interfaces as compared to *extraverts*. Orji et al. [ONDM17] showed that *conscientious* participants (a measure of carefulness or diligence) responded well to persuasive strategies such as self-monitoring and feedback in gamified systems. These studies are just a small sample of a large body of work documenting the influence of personality and cognitive ability on interactions with computer interfaces. For more detailed surveys of the literature, see [AA91, Poc91, DW96].

There is a growing interest in extending these findings to the field of data visualization [Yi12, ZOC*12a]. Some posit that knowledge of broad differences between user groups could guide the design, evaluation, or customization of systems [VHW87, ZOC*12a]. Supporting this claim, a cluster of promising research has produced evidence to suggest that individual characteristics, in addition to data mapping and visual encodings, determine the value of a visualization system. These studies have demonstrated that personality traits and cognitive abilities can have substantial impact on task performance [GF10, ZCY*11], usage patterns [BOZ*14, OYC15] and user satisfaction [Kob04]. Building on these findings, others have begun to examine how we might leverage cognitive traits for applications such as user modeling [BOZ*14, OYC15] and adaptive interfaces [LTC19].

In some circumstances, the interaction between individual differences and visualization use may have critical impact on important decision-making processes. Ottley et al. [OPH*15] investigated the impact of visualization on medical decision-making, and found that approximately 50% of the studied population were unsupported by commonly-used visualization tools when making decisions about their medical treatment. Specifically, their study showed that visual aides tended to be most beneficial for people with high *spatial ability*, while those with low *spatial ability* had difficulty interpreting and analyzing the underlying medical data when they were presented with visual representations. Another study by Conati and Maclaren [CM08] found that participants with high *perceptual speed* were less accurate in computing derived values when using radar graphs instead of heatmapped tables for data analysis. A series of studies have shown that *locus of control* (a measure of perceived control over external events) mediates search performance on hierarchical visualizations [GJF10, GF12, ZCY*11, ZOC*12b, OYC15, OCZC15]. These findings underscore the importance of incorporating individual differences into the design pipeline in order to create visualization tools that are broadly usable.

Unlike in human-computer interaction, to date there exists no comprehensive report that surveys the relevant literature on the role of individual differences in the data visualization domain. This makes it difficult to understand the scope of existing research on individual differences in this discipline, as there is no central resource researchers can consult to learn what individual differences, visualizations and tasks have been studied, and whether the results of those studies have been independently replicated. More importantly, there is limited information about how each existing study contributes to the ultimate goal of designing flexible data visualization tools that better support individual users.

In this STAR, we aim to produce a comprehensive survey that reviews the literature relevant to this topic. We identify and taxonomize existing scholarship to provide a complete picture of the current state of research, and identify possible avenues for investigation that builds upon this existing body of work. We begin by describing the scope of our review and methodology. We then proceed to a detailed analysis of the findings of this body of work. Finally, we reflect on our review to discuss core topics and opportunities for future development in this emerging area.

## 2. Existing Perspectives on Individual Differences in Visualization

The sampling of scholarly work in the previous section demonstrates the wide variety of individual differences that may be relevant to the visualization community. Pioneering work by Peck et al. [PYH*12] proposed the Individual Cognitive Differences ($ICD^3$) model which classified the space of individual differences into three dimensions (see Figure 1):

- *Cognitive traits* are the relatively stable characteristics of an individual that include features of a person's personality alongside their cognitive abilities, such as *perceptual speed*, *spatial ability*, and *visual memory*.
- *Cognitive states* are temporary mental states such as *cognitive load* and emotion. They are, by definition, transient and related to recent stimuli and the surrounding environment.
- *Experience* is the long-term construction of knowledge through exposure to real-world stimuli. *Bias* describes the predispositions one has such that one behaves in certain ways when performing certain tasks. Together, *experience* and *bias* represent a dimension that describes the accumulation of experiences that influence behavior when encountering a familiar stimulus.

Efforts to systematically evaluate *visualization literacy* (a measure of visualization experience for non-experts) [ARC*17, BRBF14, BMBH16, DJS*09, LKK16] postdate the $ICD^3$ model, but this can be viewed as a specific domain of *familiarity*.

In this STAR, we restrict the scope of our survey to focus only on the **invariant characteristics that distinguish one person from another**. Unlike *cognitive states* and measures of *experience*, the *cognitive traits* covered in this survey are believed to be stable throughout adulthood. This makes it tractable to reason about how the community can begin to incorporate individual difference into design and evaluation pipelines. Our goal is to advocate for the advancement of individual difference research in the visualization discipline by highlighting the pioneering work in this domain.

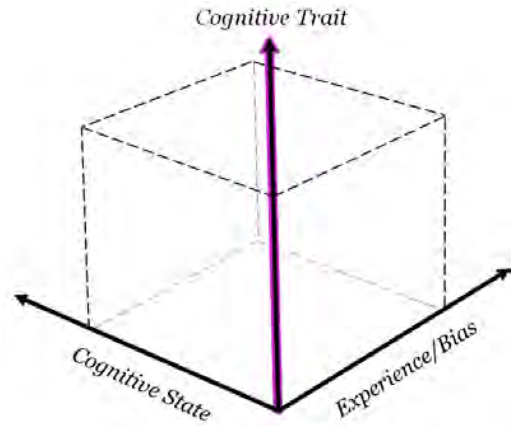

**Figure 1:** *The $ICD^3$ model from Peck et al. [PYH*12] categorizes individual differences into three orthogonal dimensions: cognitive traits, cognitive states, and experience/bias. In this STAR, we focus exclusively on cognitive trails.*

## 3. Survey Scope and Methodology

This STAR report surveys the ongoing research that studies the impact of individual differences on the use of data visualizations. The candidate papers are obtained via three methods. First, we obtain the main corpus by reviewing all the papers published on leading conferences and journals in Visualization and HCI, including InfoVis, VAST, SciVis, EuroVis, TVCG, CHI and IUI from 2008 to 2020. For this initial set of seed papers, we limit the scope to papers published in Computer Science venues (e.g., we do not collect publications from PubMed, a search engine for biomedical and life science references). Second, we search Google Scholar, ACM Digital Library, and IEEE Digital libraries with keywords such as *individual differences*, *personality*, *cognitive ability* and filter the returned results to retrieve only data visualization publications. We also web-scraped ACM Digital Library and IEEE Digital Libraries to programmatically aid the process. Finally, we followed the citations of the candidate papers obtained in the first two methods to expand the scope of our seed paper to include relevant publications that which were not published in computer science venues or were published before 2008. For all candidate papers we have collected, we manually review the title, abstract, introduction and conclusions to determine whether they are within our proposed scope. If in doubt, we also review the main content of a paper to determine its inclusion or exclusion. Finally, we removed duplicates manuscripts that studies that the same dataset. For example, [ZOC*12b] is a journal extension of [ZCY*11], so we excluded the latter. Eventually, we have found 29 key publications that are within the review scope for our main analysis.

### 3.1. Coding

We compiled a corpus of relevant literature and organized the prior work based on the *types of individual differences*, the *visualizations* used in the studies, and experimental designs such as the *tasks* and *measures* used in the experiments. During the first round of coding, a single author thoroughly read all papers to create an initial set of keywords. A second author then re-read the papers and added or consolidated the keywords when there were gaps or redundancies. For the final round of coding, two researchers who were not involved in the previous two rounds validated the coding tags and populated Table 1. The three coding rounds were not completely independent, therefore, we do not measure coding coherence.

**Table 1:** *The 29 key articles we reviewed. The filled boxes indicate the traits (■), visualizations (■), tasks (■), and measures (■) that were in the manuscripts. We use ⊠ to indicate traits that were evaluated, but no measurable effect was reported under the studied conditions. An interactive version of the table is available at* <https://washuvis.github.io/personalitySTAR>.

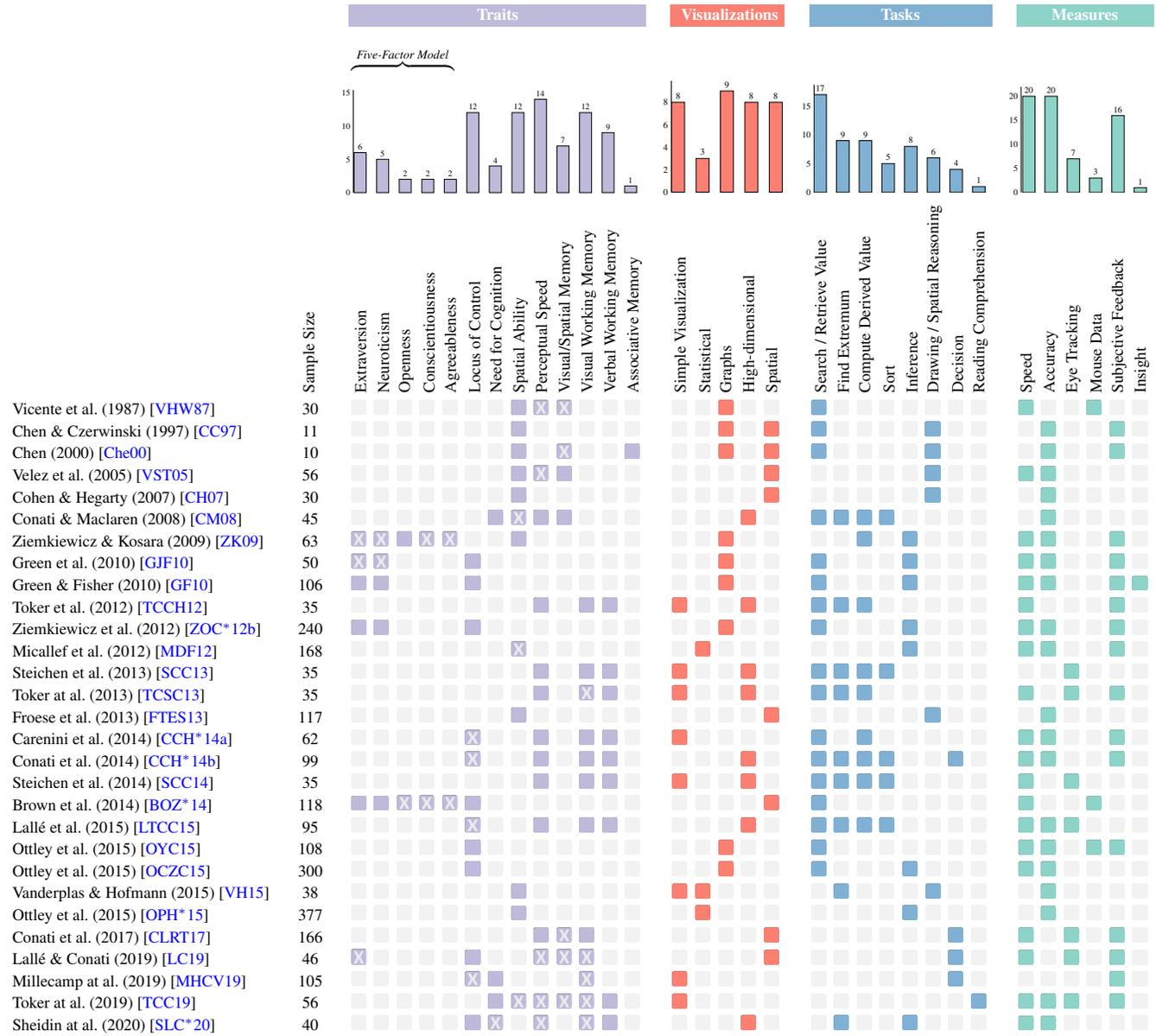

## 4. Overview of Paper

The proposed taxonomy of the publications consists of four dimensions: (1) the **Individual differences/traits** studied; (2) the types of **visualization** used; (3) the **tasks** involved in the associated experiment; and (4) the **measures** (or dependent variables) that were evaluated. Table 2 summarizes the 11 primary traits observed in the literature, which are used to organize the remainder of this paper.

We classified each paper based on the dimensions in our proposed taxonomy. A natural way to accomplish this is to assign each paper one or more tags for each of the four dimensions. For example, the earliest paper in our collection by Vicente et al. [VHW87] explored how a series of traits might impact speed and navigation for hierarchical search. Thus, the tags were *Spatial Ability, Perceptual Speed, Visual Working Memory, Networks, Search/Retrieve Value, Speed, Other Qualitative*. Using these tags, we are able to systematically analyze each paper following our taxonomy as a guide, distinguishing between **Personality Traits** and **Cognitive Abilities**. We visualize the tagging results in Table 1.

## 5. Personality Traits

Personality traits are the individual differences in thinking and behaving characteristics [All37]. The literature contains numerous examples of personality traits that interact with visualization use. For instance, researchers have uncovered that *locus of control*, a measure of perceived control, is a key factor that correlates with speed, accuracy and strategy [GJF10, GF10, BOZ*14, OCZC12, OYC15, ZCY*11, ZOC*12b]. We find that almost all of the personality traits studied in the surveyed publications are either subsets of the *Five-Factor Model* or *Locus of Control*. This is not surprising because psychologists have concluded that most personality traits are subsumed by the *Five-Factor Model* [O'C02]. *Locus of Control* has also been studied extensively in various domains [WL17].

It is important to note that researchers commonly construct hypotheses about the performance of individuals with different personal characteristics based on theories and studies established in psychology. We find that, in many cases, researchers will assume a trait to be advantageous to problem-solving with visualizations if this trait has been shown to be conducive to either problem-solving, decision-making, socioeconomic advancement or educational attainment, etc. For example, *extraversion* was hypothesized to be helpful in performing visual-related tasks [GF10]. However, personality constructs are complex and interrelated, and we observe several cases in which the results are contrary to expectation.

**Table 2:** *Definitions of the cognitive traits that are common in the visualization literature.*

| | | | |
|---|---|---|---|
| **Cognitive Traits** | | | |
| **PERSONALITY TRAITS** | **Five-Factor Model [Gol93]** | **Extraversion** | The tendency to engage with the external world. |
| | | **Neuroticism** | The tendency to experience negative emotions such as stress, depression or anger. |
| | | **Openness to Experience** | The propensity to seek, appreciate, understand and use information. |
| | | **Agreeableness** | The tendency to consider the harmony among a group of individuals. |
| | | **Conscientiousness** | The propensity to control one's impulse and display self-discipline. |
| | **Locus of Control** [Rot66, Rot75, Rot90] | | The extent to which a person believes the external world is influenced by their own actions, and/or whether they have control over the outcome of events occurring around them. |
| | **Need for Cognition** [CP82] | | The tendency to engage in and enjoy activities that involve thinking. |
| **COGNITIVE ABILITIES** | **Spatial Ability** [RS13] | | The ability to generate, understand, reason and memorize spatial relations among objects. |
| | **Perceptual Speed** [EDH76] | | The rate at which an individual is able to make accurate visual comparisons between objects. |
| | **Visual / Spatial Memory** [Spe63] | | The capacity to remember the appearance, configuration, location, and/or orientation of an object. |
| | **Working Memory** [Bad92] | | The capacity to store information for immediate use. |
| | **Associative Memory** [Car74] | | The ability to recall relationships between two unrelated items. |

**Table 3:** *The summary findings from Green and Fisher [GF10]*

|  | Completion Times | Errors | Insights |
|---|---|---|---|
| **Locus of Control** | internal locus faster times | none | external locus more insights |
| **Extraversion** | more extraverted faster times | none | less extraverted more insights |
| **Neuroticism** | more neurotic faster times | none | less neurotic more insights |

## 5.1. Five-Factor Model

The five dimensions of the *Five-Factor Model* (see [Gol93]) – *extraversion*, *neuroticism* (also referred to as *emotional stability*), *openness to experience*, *conscientiousness* and *agreeableness* – are frequently studied personalities among the surveyed publications (e.g., [ZK09, GJF10, GF10, ZCY*11, ZOC*12b, BOZ*14]); 6 out of 29 of the surveyed publications investigated one or more dimensions in the *Five-Factor Model*. Some common survey instruments of the *Five-Factor Model* include: Donnellan et al.'s Mini-IPIP [DOBL06] or De Young et al.'s 10 Big-Five Aspects [DQP07], and Johnson's 120-question IPIP NEO-PI-R [Joh14].

### 5.1.1. Extraversion

*Extraversion* is defined as the tendency of an individual to engage with the external world. Extraverts are more assertive and have stronger desire for social attention, compared to the more quiet and reserved introverts [WR17]. Extraverts have been found to achieve higher socio-economic status than introverts [Gen14]. Some studies indicate a correlation between high level of *extraversion* and higher academic achievements [CP13], while others have found contradictory results [HHL11]. The studies that find a negative correlation between *extraversion* and academic achievement hypothesize that extraverts get distracted more easily than introverts [HHL11].

**Extraversion in Visualization**
A similar contradiction exists in the data visualization domain. Green et al. [GJF10] studied how *extraversion* (among others factors) impact hierarchical search. Their initial studies found no correlation between *extraversion* and task performance. A follow up study with a larger sample size (106 versus 50 in their earlier study), however, revealed that extraverted participants completed search tasks more quickly [GF10]. In contrast, introverted subjects attained more insights from the data [GF10]. Further investigations by Ziemkiewicz et al. [ZOC*12b] partially confirmed the interaction between *extraversion* and hierarchical search. Their results showed no measurable effect on interaction time, but they found that *extraversion* impacted participants' error rates. In particular, intraverted participants were more accurate in answering the questions posed by the tasks.

Altogether, the researchers found that extraverts and introverts exhibited different problem-solving approaches. The difference in problem-solving approach was a likely explanation to the various reported results of the three studies. Specifically, Ziemkiewicz et al. [ZOC*12b] stated that, compared to extraverts, introverts took more time to understand the underlying concepts and it took them longer before attempting to solve a problem. Consequently, introverts were able to attain higher accuracy than extraverts [ZOC*12b]. This also explained why Green and Fisher [GF10] reported that introverts generated more *insights* (see Table 3). *Insight* in their study was defined as anything unexpected or novel learned by the participants while completing the tasks. Researchers speculate that the extra time taken by introverts may be very helpful in using data visualizations to solve problems, especially for unfamiliar visualizations and datasets [ZOC*12b].

### 5.1.2. Neuroticism

*Neuroticism* is defined as the tendency to experience negative emotions such as stress, depression or anger. [JRSO14]. High *neuroticism* is correlated with introversion [Uzi06] and low problem-solving skills [CRE*93]. One study found that people with average levels of *neuroticism* solved problems much faster than those with either high or low levels of *neuroticism* [Far66]. Studies in the visualization community, however, contradict this finding.

**Neuroticism in Visualization**
Green and Fisher [GF10] found that more neurotic participants completed procedural tasks faster (see their summary findings in Table 3). Ziemkiewicz et al. [ZOC*12b] also found that their neurotic participants, on average, attained high accuracy on hierarchical search tasks. It turned out visualization design mediated this effect. Their finding showed that individuals who were more neurotic tended to do well in container-style layouts, while individuals who were less neurotic did better with indented-tree layouts. The two groups of researchers speculate that one or more of the following reasons might explain their findings:

- More neurotic individuals are more attentive to tasks [IMB04], which is especially helpful when dealing with unfamiliar visualizations and data [GF10, ZOC*12b].
- More neurotic individuals are more likely to feel in control and manipulate interfaces better, similar to those with internal *Locus of Control* [GF10].
- More neurotic individuals might put more pressure on themselves to perform the tasks well [ZOC*12b].
- Since high levels of *neuroticism* are related to low *emotion stability*, Ziemkiewicz et al. [ZOC*12b] claim that feeling "out of control" is advantageous when facing unfamiliar visualizations. This hypothesis contradicts the second point (the explanation provided by Green and Fisher [GF10]).
- Less neurotic participants were either unwilling or less capable of adapting to the more unfamiliar, container style layouts and so performed poorly with those visualizations [ZOC*12b].

### 5.1.3. Openness to Experience

*Openness to experience* (or "*openness*") is defined as one's propensity to seek, appreciate, understand and use information [DGP12]. Being *open to experience* can be associated with being open-minded and curious. Psychologists have found that *open to experience* is positively related to better academic achievement [HHL11, HVRT12]. A large-scale review by Jensen [Jen15] suggests that such correlations have been found in many studies. Some scientists also believe that *openness* is beneficial for professional development [NZ15] and that people who score high on *openness* tend to have higher intellectual ability [AH97, GSL04] and better problem-solving skills [MSDL15].

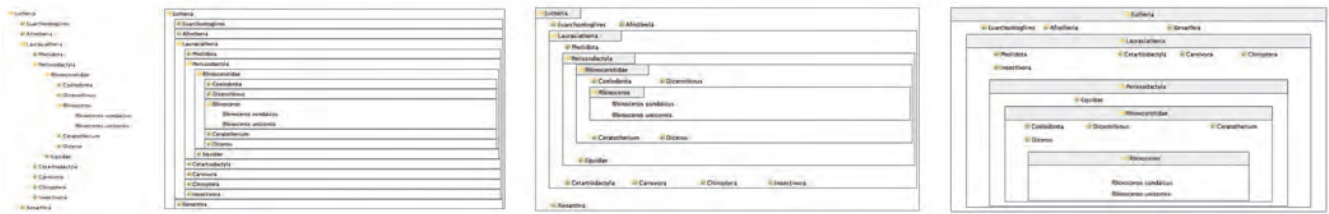

**Figure 2:** *Visualization layouts used in Ziemkiewicz et al.'s studies on the influence of the five-factor model and locus of control on hierarchical search tasks, each displaying the same data [ZCY*11, ZOC*12b].*

**Openness in Visualization**
Researchers in the data visualization domain commonly construct hypotheses based on theories and prior results in psychology. However, *openness* is largely under-explored by visualization researchers. A single study by Ziemkiewicz and Kosara [ZK09] found that participants who scored high on openness had easier time overcoming conflicting visual and verbal metaphors when solving problems related to hierarchical visualizations. Brown et al. [BOZ*14] found no measurable impact of *openness* on visual search strategies.

### 5.1.4. Conscientiousness

*Conscientiousness* is defined as the propensity to control one's impulse and display self-discipline. A high score in this dimension is associated with being focused and goal-oriented [RJF*09], but it is also related to stubbornness and being overly-demanding [LOR*11, CIV*19]. On the contrary, a low score on the *conscientiousness* scale is connected to being unreliable and lack of focus [MA03]. Overall, many researchers believe that high *conscientiousness* is related to career success [SCJME09, Tou12] and better problem-solving skills [DMOGP11].

**Conscientiousness in Visualization**
In the data visualization domain, however, *conscientiousness* has not been well-studied. In fact, to the best of our knowledge, there is no recent publication that investigates how *conscientiousness* affects the use of visualizations. A few studies [ZK09, BOZ*14] measured *conscientiousness* alongside the other five-factor traits, but found no significant impact.

### 5.1.5. Agreeableness

*Agreeableness* measures a person's tendency to consider the harmony among a group of individuals [RC03]. Conversely, disagreeableness/*low agreeableness* is associated with prioritizing one's self-interest. *Agreeableness* is considered to be a beneficial trait for performing collaborative tasks in teams [DGSO06, PVTR06]. Scoring low in *agreeableness*, however, can also be potentially advantageous because some researchers have found that low *agreeableness* is associated with creativity [KLWG13]. Also, there are contradictory opinions on whether *agreeableness* is positively related to academic achievement [LSLG03, HHL11] or not [Dis03].

**Agreeableness in Visualization**
As with to *conscientiousness*, *agreeableness* has yet to be studied in-depth by visualization researchers although it has been measured as part of the Five-Factor Model in a small number of studies [ZK09, BOZ*14]. Both Ziemkiewicz et al. [ZK09] and Brown [BOZ*14] found no effect on search tasks.

## 5.2. Locus of Control

*Locus of control* measures the extent a person feels in control of or controlled by external forces [Rot66, Rot75, Rot90]. Individuals fall on a continuous spectrum, with one end being internal *locus of control* (**Internals**) and the other end being external *locus of control* (**Externals**). The Internal-External Locus of Control Inventory is a popular measure to evaluate an individual's *locus of control* [Rot66]. Low scores are associated with internal *locus of control* and high scores are associated with external *locus of control*. According to Rotter [Rot90], individuals who exhibit internal *locus of control* believe that they have control over their own actions, the actions' outcomes and the environment around them. In contrast, those who exhibit external *locus of control* tend to attribute outcomes to external forces. *Internals* tend to be more confident [Hei10] and optimistic [BH15] than *Externals*. Researchers also believe that internal *locus of control* is associated with academic achievements [FC83, GBPM06]) and strong problem-solving skills [MR93, OS15].

### 5.2.1. Locus of Control in Visualization

Green et al.'s experiments [GJF10, GF10] were among the first to study the relationship between *locus of control* and user performance with visualization-related tasks. They conducted a study [GF10] to investigate the relationship between *locus of control* and search performance across two hierarchical visualization designs. They found that *Internals* were significantly faster than *Externals* when performing procedural tasks (search tasks to locate items). However, *locus of control* had no significant impact on accuracy. *Externals*, however, reported more *insights* than *Internals*.

Ziemkiewicz et al. [ZCY*11] extended Green et al.'s work [GJF10, GF10] to further investigate how *locus of control* affects visualization use. They hypothesized that layout (defined as the spatial representation and arrangement of visual marks in a visualization [ZCY*11]) was the determining factor in the interaction between *locus of control* and visualization usage. They further hypothesized that *Internals* would have difficulties using visualization

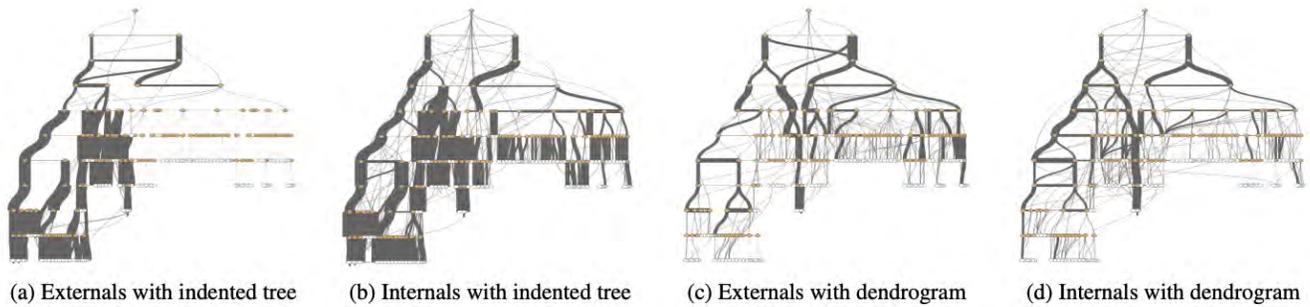

(a) Externals with indented tree     (b) Internals with indented tree     (c) Externals with dendrogram     (d) Internals with dendrogram

**Figure 3:** *Visualization of different search pattern observed in Ottley et al.'s study, grouped by locus of control (external vs. internal) as well as visual layouts [OYC15]. The thickness of the each line between every two nodes is proportional to the number of participants who explored that path.*

that were more "contained", while *Externals* would be able to adjust to various visual layouts. To test their hypotheses, Ziemkiewicz et al. [ZCY*11] designed four visualizations that differed only in layout. They designed and tested a set of visualizations that gradually transitions from an indented list layout to a containment layout, while keeping constant the interaction mechanisms (e.g., zooming v.s. scrolling), color encoding, and fonts. Figure 2 shows the four visual metaphors used by Ziemkiewicz et al. [ZCY*11, ZOC*12b].

Overall, the results of Ziemkiewicz et al. [ZCY*11]'s study showed that *Externals* were faster and more accurate than *Internals*. The performance differences were especially pronounced in the cases where participants used more "contained" visualizations (see the 3rd and 4th visual layouts in Figure 2). One interesting result was that *Internals* were significantly slower than *Externals* in completing inferential tasks [CWCO19] (such as comparing two items/objects found in the visualization), although *Internals* and *Externals* completed procedural tasks at approximately the same speed. Ziemkiewicz et al. [ZCY*11] speculated that *Externals* were better than *Internals* at adapting their thinking to external representations (such as the layout of a visualization) since they were more inclined to rely on external conditions rather their own internal representations and processes.

Although *locus of control* is believed to be relatively stable throughout adulthood, psychologists have found that it is possible to temporarily influence a person's *locus of control* score [JGPC92, FJ96]. Some researchers see this as an opportunity to resolve design challenges. Further investigations by Ottley et al. [OCZC15] replicated Ziemkiewicz et al.'s [ZCY*11] experiment design to study whether changes in *locus of control* can predictably influence performance. The priming method used in their study was based on Fisher and Johnson's technique [FJ96]. This technique works by asking a person to recall examples of times when they feel either in control of (priming *Externals* to be more *internal*) or out of control of (priming *Internals* to be more *external*) the situations. The results of Ottley et al. showed that priming was largely effective [OCZC15]. For example, when *Internals* were primed to be more *external*, they exhibited performance measures similar to participants grouped as *Average* by Ziemkiewicz et al. [ZCY*11]. Similarly, *Average* participants who were primed to be more *inter-*

*nal* produced performance measures similar to the *Internals* of the control group. The only exception was *Average* primed to be *external*. Their behaviors differed from the control group.

In addition to these performance differences, researchers believe that it is also possible for *locus of control* to affect behavioural patterns [OYC15]. To investigate this, Ottley et al. [OYC15] analyzed the strategies employed by *Externals* and *Internals* with two different hierarchical visualizations (indented trees and dendrograms). For indented trees, *Externals* followed the top-down design of the indented tree and adopted a strategy similar to depth-first search, while *Internals* followed a strategy that somewhat resembled breadth-first search. For dendrograms, *Externals* were more sporadic when they navigate the visualization, while *Internals* pursued a combined depth-first search and breadth-first search strategy. Figure 3 shows the various strategies observed (note that the thickness of a route is proportional to the number of participants observed to follow that path). The results showed that *Externals* performed better (they found the targets faster) with indented trees, while *Internal* were superior with the dendrogram. Similarly, Brown et al. [BOZ*14] found that *Internals* and *Externals* applied different searching strategies when performing a visual search task.

### 5.3. Need for Cognition

Cohen et al. first described *need for cognition* in 1955 as "a need to structure relevant situations in meaningful, integrated ways. It is a need to understand and make reasonable the experiential world." [CSW55]. In more recent conceptualization, the term has come to mean a "chronic tendency to engage in and enjoy effortful activities" [CPFJ96], such as reading and solving puzzles.

According to Cacioppo and Petty's characterization of this concept [CP82], individuals with high *need for cognition* are more likely to make sense of their world by seeking, acquiring, and reflecting on information. In contrast, those with low *need for cognition* are more likely to rely on others (e.g., experts and famous people), heuristics, or social comparisons to make meaning of events, relationships, and other stimuli. One common tool for assessing need for cognition is a 34-item instrument developed by Cacioppo and Petty [CP82], which scores participants along a continuum from low to high *need for cognition*. A later version condensed the number of items to 18, with no appreciable loss of discriminatory power [CPFK84], and this short form is the most common tool used to measure *need for cognition* in visualization-related studies, e.g. [CM08, MHCV19, TCC19].

Several studies have sought to evaluate the correlation between *need for cognition* and other measures of individual difference (see [CPFJ96] for a complete survey). Amabile at al. observed a significant positive correlation between *need for cognition* and intrinsic motivation, as well as a corresponding negative correlation with extrinsic motivation [AHHT94]. Fletcher at. al found that people with a higher *need for cognition* tended to have a significantly more internal *locus of control* [FDF*86]. Cacioppo et al. have suggested the existence of a relationship between *need for cognition* and the *conscientiousness* and *openness* dimensions of the Five-Factor Model [CPFJ96], but as of this writing, this link has not been experimentally validated.

## 5.4. Need for Cognition in Visualization

In an early investigation of the effect of *need for cognition* in visualization, Conati and Maclaren [CM08] conducted a study to evaluate the efficacy of various individual differences (including *need for cognition*) in predicting the relative effectiveness of a radar graph and a heatmap for various tasks. They found that in conjunction with other measures, *need for cognition* had a positive relationship with accuracy in sorting tasks using the heatmap . They also found that this relationship was not present in trials utilizing the radar chart. However, the authors note that while they did observe a statistically significant effect, the models explain only a small proportion of the overall variance, suggesting that the effects of *need for cognition* are likely moderated by other, yet unobserved features.

Millecamp et al. found that *need for cognition* plays a role in a person's response to visual explanation of recommendations in a music recommender system [MHCV19]. Using a custom recommendation interface built on top of Spotify (see Fig. 4), the study varied whether or not participants interacted with a baseline system or with an augmented version including both bar chart and scatterplot views containing more information regarding why a selected song was recommended. They observed a statistically significant interaction effect between *need for cognition* and the participants' subjective ratings of confidence. Specifically, there was a modest increase in confidence for participants with *low need for cognition* in the visual explanation condition compared with the baseline, and a modest decreased in confidence for participants with *high need for cognition* in the visual explanation condition compared with the baseline.

Toker et al. [TCC19] observed that *need for cognition* had a significant positive effect on participants' accuracy when performing recall tasks with a bar chart as a component of a Magazine-Style Narrative Visualization, but that there was no statistically significant relationship to speed. This may at first seem counterintuitive: one would expect that participants with higher *need for cognition* would be able to perform more quickly, and that their commitment to synthesizing all available information would improve their accuracy. Upon closer inspection, however, we observe that in this study, the term *speed* refers to the total time spent interacting with the visualization. When this meaning is applied, the positive relationship between *need for cognition* and time spent interacting with a visualization are in line with observations made in non-visualization contexts: because people with higher *need for cognition* are predisposed to engage in sensemaking behavior, it makes sense that they would spend more time trying to understand the visualization before moving on to the subsequent task. However, these findings were inconsistent with a followup study by the same authors [TCC19], wherein they reported no significant relationship to time on task but did observe a relationship with accuracy. Additionally, this latter study included an analysis of eye-tracking data, but found no significant relationship [TCC19]. These conflicting results suggest that more investigation is needed into the role of *need for cognition* in visualization use.

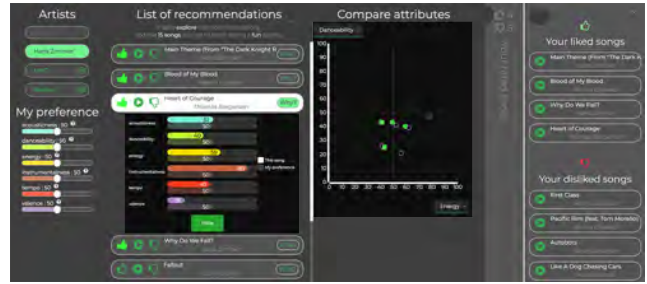

**Figure 4:** *The music recommender interface from Millecamp et. al's 2019 study on the effects of need for cognition on participants' response to visual explanation of recommendations. The interfaces for the control condition differed from the stimulus condition only by the omission of the two highlighted regions, which provide explanations about why a song was recommended.*

## 6. Cognitive Abilities

*Cognitive abilities* refer to mental capabilities in problem solving and reasoning (including *visual reasoning*) [IB15]. The data visualization community has extrapolated the effects of *cognitive abilities* on the users' performances and experiences with visualizations from foundational research in psychology. We find literature related to *spatial ability* [CC97, Che00, VST05, ZK09, FTES13, OPH*15, VH15], *perceptual speed* [CM08, TCCH12, TCSC13, SCC13, CCH*14a, CCH*14b, SCC14], *visual working memory* [CM08, DMBM09, APM*11, TCCH12, SCC13, TCSC13, CCH*14b, SCC14], *verbal working memory* [TCCH12, TCSC13, SCC13, CCH*14a, SCC14], and *associative memory* [Che00].

## 6.1. Spatial Ability

*Spatial ability* is broadly defined as the capacity to generate, understand, reason and memorize spatial relations among objects [RS13]. Though there is no consensus on precisely which mental abilities are encompassed by this general term, commonly referenced components include *spatial orientation, spatial location memory, targeting, spatial visualization, disembedding* and *spatial perception* (for further detail on these concepts, please see [Kim00]). Individuals with high *spatial ability* tend to excel in scientific and engineering fields [WLB09] and exhibit stronger problem-solving skills for various tasks [WHA*02, CKT16, SDL*18], including solving mathematical [YLM18] and geometric problems [BSC19]. Some commonly used tests for *spatial ability* include the *paper folding test* [EDH76] and *mental rotation test* [VK78].

### 6.1.1. Spatial Ability in Visualization

Given the importance of *spatial ability* in analytical contexts, the relationship between this construct and visualization use has generated substantial interest in the visualization research community. Early work by Vicente et al. [VHW87] investigated how *spatial ability* influenced interactions with computer-based visualizations. In this study, participants were asked to locate a piece of information in a hierarchical file system. The researchers found that *spatial ability* was a significant predictor of completion time, and they concluded that *spatial ability* had a dramatic impact on performance. Later studies found the *spatial ability*'s influence on visualization use and performance might not be as straightforward as one would expect. In the information retrieval domain, Chen and Czerwinski [CC97] reported that *spatial ability* was positively correlated with recall, but negatively correlated with precision, and these findings were partially replicated in a follow up study [Che00].

Most studies, however, have consistently reported that *spatial ability* is positively correlated with performance in various visual tasks. For example, Velez et al. [VST05] found that participants with higher *spatial ability* were faster and more accurate at identifying real and computer-generated 3-D objects when given the objects' orthogonal projections from various perspectives. Cohen and Hegarty [CH07] asked participants to sketch the cross section of a computer generated 3-D object, and observed that individuals with higher *spatial ability* generally performed better, and that these same participants were more likely to make use of supporting animation.

As in their investigation of *openness*, Ziemkiewicz and Kosara [ZK09] observed that individuals with high *spatial ability* were better equipped to overcome incompatible visual and verbal metaphors when navigating hierarchical data. In their study on the efficacy of visualization and structured text in supporting medical decision-making, Ottley et al. [OPH*15] reported that participants with higher *spatial ability* were more accurate and faster than the group with low *spatial ability*, and were better able to make use of the more text+visualization representation of the data. Similar performance advantages were reported in VanderPlas and Hofmann's [VH15] experiment with lineup tasks, and Conati and Maclaren [CM08] found that *spatial ability* was positively correlated with better performance in characterizing distributions. However, Froese et al. [FTES13] found that people with low *spatial ability* experienced significant performance gains after being trained in using visualizations.

These studies broadly suggest that *spatial ability* has a largely positive relationship with performance when using visualizations. One possible explanation is that *spatial ability* might affect a participant's strategy or usage pattern. For example, Vicente et al. [VHW87] found that individuals with low *spatial ability* frequently descended an incorrect path through the hierarchy, requiring them to backtrack. Chen and Czerwinski [CC97] observed that participants with high *spatial ability* commonly combined detailed local moves with strategic jumps that exploited the global structure of the visualization, whereas those with low *spatial ability* tended to remain at the local level.

### 6.2. Perceptual Speed

*Perceptual speed* measures the rate at which an individual can scan and compare figures and symbols, as well as perform simple visual perception tasks [EDH76]. Studies have demonstrated links between high *perceptual speed* and educational achievement [Mel82], information retrieval [All92], and acquiring programming skills [Shu91]. Some commonly used tests for *perceptual speed* are the *Identical Pictures Test* [EDH76], the *Finding A's Test* [EDH76] and *Number Comparison Test* [EDH76].

### 6.2.1. Perceptual Speed in Visualization

Vicente et al.'s [VHW87] pioneering study on *spatial ability* included *perceptual speed* as one of the candidate predictors of user performance. However, they found no measurable effect of *perceptual speed* on searching hierarchical file systems. More recent investigations by Conati and Maclaren [CM08] found that *perceptual speed* mediate tasks performance. For example, they found *perceptual speed* to be positively correlated with the accuracy of "computing derived values", a category of tasks defined by Amar et al. [AES05] that involves deriving an aggregate number from graphical data. Overall, the found that participants with low *perceptual speed* did better than those with high *perceptual speed* with radar graph, while the opposite was true for heatmapped tables [Wil04] (see Figure 5). Toker et al. [TCC12] also found that individuals with high *perceptual speed* completed tasks faster with both radar and bar graphs. Similar results were reported by [CCH*14a, CCH*14b, LCC17].

One study also reported that high *perceptual speed* led to higher learning rate (measured by the change in task completion time or accuracy over time) [LTCC15]. Toker et al. [TCC19] found that individuals with low *perceptual speed* had difficulties remembering legend details and axis labels. Further studies by Toker et al. [TCSC13], Steichen et al. [SCC13] and Conati et al. [CLRT17] all showed that it was possible to infer a user's *perceptual speed* dynamically based on eye-tracking data.

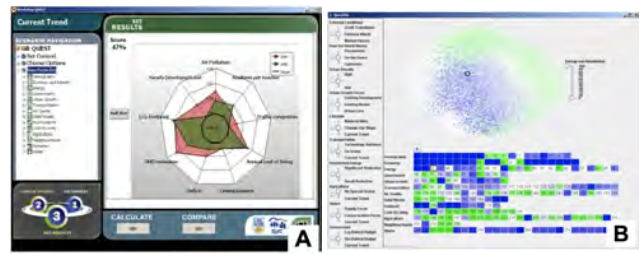

**Figure 5:** *The radar graph (A) and heatmapped tables (B) used by Conati and Maclaren in their exploration of the relationship between perceptual speed and task performance [CM08].*

### 6.3. Visual / Spatial (Short-Term) Memory

*Visual / spatial memory* measures the short-term ability to remember the configuration, location, and orientation of an object [Spe63], and is commonly measured using Eckstrom et al.'s Shape Memory Test (MV-1) [EDH76] or other instruments. The visuospatial nature of data visualization suggests an intuitive link between an individual's *visual memory* and their performance using visualization tools, and this intuition has led to an abundance of studies investigating this relationship. However, the results of these investigations have been mixed.

#### 6.3.1. Visual / Spatial Memory in Visualization

Several of the studies described in Sections 6.1 and 6.2 also investigated the role of visual memory. In Vicente et al.'s 1987 study found no relationship between *visual memory* and how people navigate hierarchical file systems [VHW87]. Similarly, Chen's 2000 study [Che00] ( see section 6.1) found no relationship between *visual memory* and search performance in a spatial-semantic virtual environment. Velez et al. [VST05] did observe a statistically-significant relationship between *visual memory* and accuracy in their projection task, but the influence was modest. Conati and Maclaren [CM08] reported a similar relationship during filter tasks using the heatmapped table, but not the radar chart.

Participants with low *spatial memory* in Lallé at al.'s study of user experience reported that they found the MetroQuest interface substantially less useful [LCC17, LC19]. In a companion analysis to this study, Conati et al. [CLRT17] found that eye tracking data could be used to accurately predict participants' *spatial memory*, suggesting that this feature is associated with distinct gaze patterns in visualization use.

### 6.4. Working Memory

Many of the studies that investigated *perceptual speed* also evaluated *working memory*, a measure of an individual's capacity for temporarily storing and manipulating conscious perceptual and linguistic information [Bad92, MS99]. This term was originally coined in 1960 by Miller et al. in the context of their work on theory of mind [MGP60], and is distinct from *short-term memory* in that the emphasis is on the **active manipulation** of information, rather than simple recall [Cow08].

Daneman and Carpenter first observed a link between *working memory* and *reading comprehension* [DC80], and this relationship has been independently verified by many other studies [DM96]. It appears to play a substantial role in academic achievement [SBF04, AA10], as well as in attention [FV09], though the latter finding has been recently called into question following a more nuanced investigation using eye tracking [MMWL14]. *Working memory* is of particular interest to visualization researchers because of its significance in supporting reasoning [Voo97, Kla97, CHD03], decision-making [HJW02, HJW03, Brö03], and other cognitive processes critical to effective analysis [Dia13].

#### 6.4.1. Working Memory in Visualization

Two different forms of working memory are frequently assessed in the visualization literature. *Visual working memory* is a mecha-nism by which visual information (including position, shape, color, and texture) is retained between eye fixations [LV97]. This enables cognitive actions such as change detection [LV13]. *Verbal working memory* is responsible for temporarily storing and manipulating language-related information, including both words and numerical values [vDM16]. This enables actions such as remembering a telephone number long enough to dial it [MD16]. A commonly cited test for *visual working memory* is a set of change detection tasks of colored squares developed by Edward K. Vogel and collaborators ( [LV97, VWL01, FV09]). For measuring *verbal working memory*, Operation-Word Span Test (OSPAN) [TE89] and the Corsi Test [Cor72] are found in the surveyed literature.

Toker et al. [TCCH12] found a statistically-significant, divergent relationship between participants' *working memory* and their preference ratings of bar charts and radar plots. Specifically, participants with higher *visual working memory* rated radar graphs as more preferable, and those with lower *verbal working memory* tended to rate bar graphs as easier to use. In follow-ups to this study using the same interfaces and tasks, Steichen et al. [SCC13, SCC14] found that eye tracking data could be used to accurately predict both *visual* and *verbal working memory*. However, further analysis found that only *verbal working memory* was statistically significant in the prediction of specific gaze behaviors [TCSC13].

In their investigation of the effects of highlighting interventions on speed and accuracy on search and comparison tasks using bar charts, Carenini et al. [CCH\*14a] found that participants with low *visual* and/or *verbal working memory* consistently underperformed on comparison tasks. This effect was absent for simple search tasks. Conati et al. observed similar relationships in their study of Value Charts [CCH\*14b], and further demonstrated that layout appears to partially mitigate this performance deficit. Further analysis by Lallé et al. [LTCC15] found both forms of *working memory* useful in predicting participants' learning curve on this interface, characterized by the rate of change in response time over multiple trials.

In their experiments on the MetroQuest system, Lallé at al. [LCC17] observed a relationship between *visual working memory* and both user preference and gaze behavior. Specifically, participants with higher *visual working memory* tended to prefer charts over maps, and correspondingly tended to have more fixations on the chart areas. In these experiments, there was no relationship observed between *working memory* and willingness to utilize available interface customization options [LC19]. As with *spatial memory*, a deeper analysis of the gaze data from this experiment by Conati et al. (2017) [CLRT17] demonstrated that gaze data can be used to accurately predict participants' *visual working memory*.

In Toker et al.'s experiments on Magazine Style Narrative Visualizations [TCC19], *verbal working memory* was observed to have an intuitive negative correlation with time on tasks. They did not, however, observe any statistically significant correlation with accuracy, understanding, or interest. While *visual working memory* was measured in participants of Millecamp et al.'s experiments involving a more music curation task, they also did not observe any statistically significant relationship with this feature.

## 6.5. Associative Memory

*Associative memory* refers to a person's ability to recall a relationships between two unrelated items (for example, linking a name and a face) [Car74]. It can be measured by MA-1 scores [EDH76]. Some researchers believe that it is valuable to investigate the effects of *associative memory* on user interaction with data visualizations, because good *associative memory* helps building mental maps of virtual environments or interfaces and can aid users in navigating the virtual spaces [Che00].

### 6.5.1. Associative Memory in Visualization

To the best of our knowledge, Chen [Che00] is the only publication to investigate *associative memory* in the existing data visualization literature. In Chen's study, participants used an interactive graph of published articles and were asked to retrieve as many papers as possible for a given topic within 15 minutes. Chen found that *associative memory* was positively correlated with people's ability to retrieve the appropriate papers. Chen also reported the subjective feedback of users and found that those with good *associative memory* were more likely to believe that the spatial interface was useful.

## 7. Discussion of Findings

Although our organization of the literature on individual differences in visualization is intended to provide a broad overview of existing work in this area, we acknowledge that any post-hoc categorization (such as the traits, visualizations, tasks, and measures reported in this STAR) will not be exhaustive. Despite this fundamental limitation, our taxonomy enabled several useful insights regarding this body of work. Foremost among them were two important takeaways:

1. With very few exceptions (namely, *conscientiousness* and *agreeableness*), **there is evidence that nearly every cognitive trait in Table 2 can impact visualization use**. This body of work underscores that designing and evaluating tools to help people think is a complicated endeavor.

2. Despite the breadth of cognitive traits under investigation, there have been a relatively small number of studies which at times yielded conflicting findings. Further investigations, **including replication studies**, are crucial to enriching our understanding how individual differences impact visualization use, and to subsequently develop guidelines for the integration of this knowledge into the design of future systems.

In the following sections, we expand upon these observations in the context of several different dimensions of our taxonomy.

### 7.1. Traits

The impact of some individual differences are clear, having been replicated under multiple experimental conditions by two or more independent researchers. One such example is the consistent demonstration that *locus of control* impacts speed and accuracy on hierarchical search tasks [GJF10, ZOC*12b, OCZC12, OYC15]. This has been replicated by several studies which used comparable datasets, tasks, and measures, and the results appear to hold for both in-person laboratory experiments [GJF10] and crowdsourced studies [ZOC*12b]. Furthermore, these studies suggest that the design of the visualization itself is a significant factor [ZOC*12b], and that *locus of control* influences search strategy [OYC15].

It is interesting to note that *verbal working memory* is the only trait that has reliably resulted in statistically significant findings. *Verbal working memory* is believed to affect the processing of verbal component of visualizations, such as labels, legends, description of tasks, and texts [TCCH12, TCSC13, TCC19]. In particular, high *verbal working memory* users spend less time reading and processing various textual information in visualizations [SCC13]. An analysis of eye tracking data by Toker et al. [TCSC13] indicated that participants with low *verbal working memory* referred back to task question descriptions more frequently, and tended to scan between different parts of the screen more frequently than their high *verbal working memory* counterparts. Another study found that *verbal working memory* was positively correlated with learning rate [LTCC15]. Overall, studies have consistently reported an inverse correlation between task completion time and *verbal working memory*, though we hesitate to generalize these findings to realworld scenarios. This correlation may be attributable to unintended situational effects of the design of traditional user studies, which explicitly require participants to process textual information when completing visualization-related tasks.

Results are more ambiguous for most other traits that have been studied due to the lack of replication studies. For example, although a series of manuscripts report that *perceptual speed* impacts visualization use (9 out of 29 papers report significant effects), they inspected a range of visualization designs, tasks, and measures, making it challenging to uncover general patterns. A similar phenomenon exists for visual working memory. Although every paper in this STAR used established psychometric batteries from the psychology field, inconsistency among the surveys used to assess traits also makes it difficult to compare findings between studies. For example, both Ottley et al. [OPH*15] and Micallef et al. [MDF12] have investigated the impact *spatial ability* on Bayesian inference with visualization, but reported contradictory findings. Both studies used the same paper folding task to assess *spatial ability* [EDH76], but differed in the application of the assessment instrument: Micallef et al. [MDF12] used 10 out of 20 questions in the scale, while Ottley et al. [OPH*15] used all 20 questions. Such inconsistencies underscore both the importance of replication and the need to standardize the instruments used to assess both individual differences and task performance.

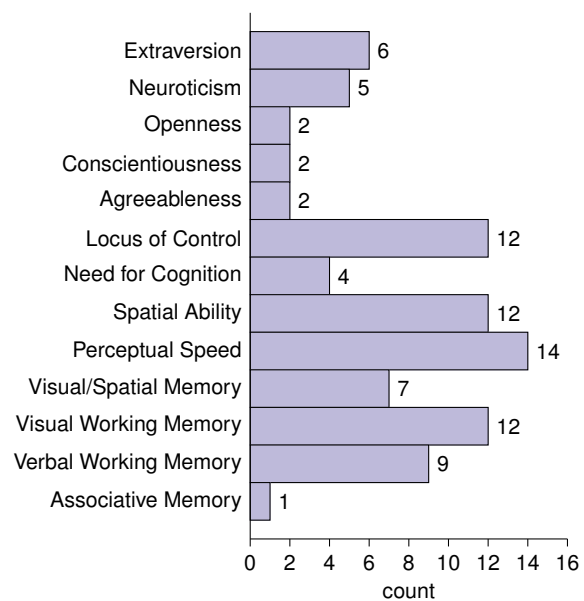

**Figure 6:** *The types and distribution of traits that were investigated in the literature on individual differences in visualization use.*

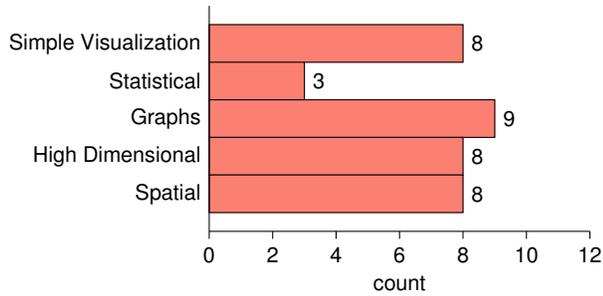

**Figure 7:** *The types and distribution of visualization designs observed in the literature on individual differences in visualization use.*

Other traits remain underexplored despite promising initial findings. As mentioned in Section 6.5, Chen [Che00]'s singular study on *associative memory* showed a positive correlation between this trait and performance on a graph navigation tasks. Similarly, Ziemkiewicz and Kosara [ZK09] found that *openness to experience* predicted easier adjustment to disruptions in visualization interaction, an observation which has promising implications for visualization scenarios involving unfamiliar or novel designs. Other traits such as *conscientiousness* are also sparsely explored in the context of visualization use, with only 2 of 29 manuscripts inspecting this trait. Both studies reported null results, though it is impossible to draw comparisons between these studies due to their vastly different experimental designs.

### 7.2. Visualization

We observed five categories of visualization design in the surveyed literature: Simple Visualization, Statistical, Graphs, High-Dimensional, and Spatial. *Graphs* were the most commonly tested visualization in the individual differences literature, appearing in 9 out of 29 surveyed papers. We observed substantial variance in the choice of both encoding and aesthetic design. The research exploring hierarchical visualization has largely focused on the impact of *locus of control* [GF12, OYC15, OCZC12, ZCY*11, ZCY*11] and the *five factor model* [GF12, ZCY*11, ZCY*11, ZK09]. Several studies also report that search and navigation with graphs and trees is influenced by *spatial ability* [Che00, CC97, VHW87, ZK09].

*Simple data visualizations* were also relatively common in the literature (8 out of 29 papers surveyed). For example, Toker et al. [TCSC13] and Steichen et al. [SCC13] found that *perceptual speed, visual working memory*, and *verbal working memory* can influence how people deploy attention to visual elements within grouped bar charts. VanderPlas and Hofmann [VH15] used histograms and dotplots among other charts, and found that *spatial ability* correlated with performance when identifying which plot was "the most different" in a collection.

VanderPlas and Hofmann [VH15] also included *statistical plots* such as boxplots, violin plots and QQ-plots, and found a similar correlation between performance and *spatial ability*. Micallef et al. [MDF12] and Ottley et al. [OPH*15] investigated *statistical plots* such as icon arrays (also known as frequency grids) and Euler diagrams. As reported in Section 7.1, these studies reported contradictory results on whether or not *spatial ability* influenced performance on Bayesian inference tasks.

A series of studies investigated individual differences in the context of radar plots (e.g., [CM08], [SCC13], and [TCSC13]). A later study by Sheidin et al. [SLC*20] compared speed and accuracy across a variety of tasks with different time series visualizations, including line charts, stream graphs, radar charts, and circle charts. They found a significant interaction between *locus of control* and speed and accuracy in some task types, and observed that *verbal working memory* also influenced completion times. Taken together, these findings suggest a correlation between *perceptual speed, visual working memory*, and *verbal working memory* and visualization use.

Studies of *spatial ability* in the context of *spatial visualization* universally reported significant effects [CC97, Che00, CH07, FTES13, VST05]. For example, Chen et al. [Che00] observed that *spatial ability* was correlated with graph search performance in virtual environments. Froese et al. [FTES13] demonstrated that training programs for creating projections of 3D objects were most beneficial for participants with *low spatial ability*.

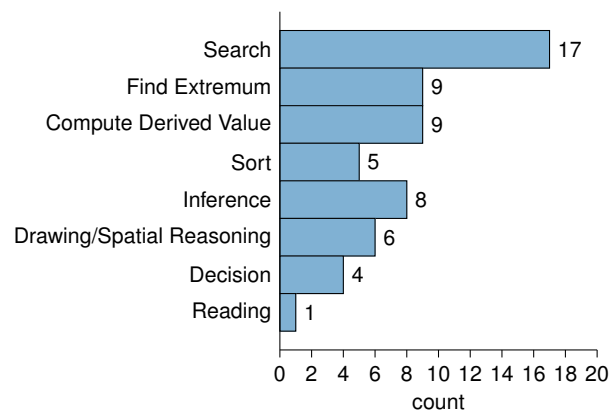

**Figure 8:** *The types and distribution of tasks observed in the literature on individual differences in visualization use.*

## 7.3. Tasks

We observed a wide variety of tasks in the literature, many of which are based loosely on Amar et al.'s analytic task taxonomy [AES05] (e.g., *information retrieval/search*, *find extremum*, *compute derived values*, and *sort*). *Search* was the most common type of task that we observed in the literature. Examples range from finding documents in a file structure [Che00, VHW87] or phylogenic tree [GF10,OCZC15,ZOC*12b] to finding Waldo [BOZ*14].

Other tasks were less common. For example, Chen & Czerwinski [CC97] asked their experiment participants to draw the "semantic space" (a term describing the user interface, which consists of nodes and links among them) from memory. They found that those with higher *spatial ability* were more accurate in recalling the visualization structure. Toker et al. [TCC19] asked subjects to use a "Magazine Style Narrative Visualization" that supplemented textual documents and answer reading comprehension questions. Millecamp et al. [MHCV19] asked participants to use a custom interface (Fig. 4) to create music playlists for different activities.

## 7.4. Measures

The majority of existing studies assessed the effects of personality traits and cognitive abilities through traditional measures of performance such as speed (20 out of 29 papers) and accuracy (20 out of 29 papers). With a few exceptions (e.g. [CM08]), high scores in the studied *cognitive abilities* were correlated with better task performance. For example, higher levels of *spatial ability* correlated with better statistical reasoning [OPH*15], and high *perceptual speed* predicted superior ability to find similarities and differences among objects [AI94].

In contrast, the results for personality traits are more nuanced, with effects that are moderated more significantly by visualization design and task. For example, Ottley et al. [OYC15] compared search speed across two tree visualization designs: a dendrogram and an indented tree. They found that participants with *external locus of control* were faster than their *internal locus of control* counterparts when performing search tasks with an indented tree visualization. However, they also observed the reverse when they analyzed interaction times for the dendrogram, suggesting that neither of the designs were suitable for both groups of users.

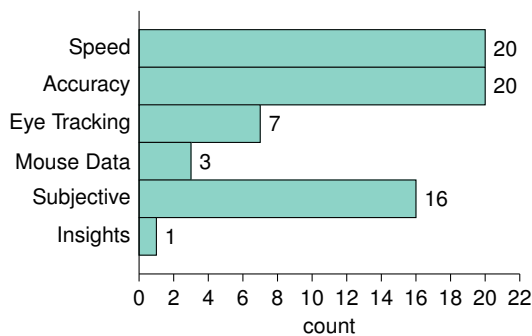

**Figure 9:** *The types and distribution of measures observed in the literature on individual differences in visualization use.*

In addition to speed and accuracy, studies frequently solicited subjective feedback to evaluate users' experiences [GJF10, TCCH12, LCC17, LC19, TCC19]. For instance, Ziemkiewicz et al. [ZOC*12b] captured participants' preference for the designs in their study. In Micallef et al.'s [MDF12] study on statistical inference, subjects recorded their confidence in the answers given. Chen [Che00] found subjects with strong *associative memory* were more likely to believe that the graph-based visualization interface was useful. Toker et al. [TCCH12] found that participants with higher *visual working memory* preferred radar charts more than those whose *visual working memory* was lower.

Eye-tracking has seen significant use in the surveyed literature (7 out of 29 papers). Beyond measures of speed and accuracy, eye movements can provide important information about how different representations facilitate information processing. Eye gaze data has been used to predict both task and visualization type [SCC13], as well as task complexity as defined by the study [SCC14], user performance [SCC14, TCC19], and learning curve [LTCC15]. In addition, one study found that *perceptual speed* was positively correlated with more efficient visual scanning behavior [TCSC13].

Three of the surveyed studies used mouse interaction to explore how individual differences impact visualization use. Vicente et al. [VHW87] found that *spatial ability* correlated with scrolling behavior with a hierarchical file structure. Ottley et al. [OYC15] showed that different *locus of control* groups exhibited distinct patterns of mouse movement when searching hierarchical visualizations. Finally, Brown et al. [BOZ*14] used machine learning to predict *locus of control*, *extraversion*, and *neuroticism* from features that were derived from mouse clicks and mouse movements during a visual search task.

A single study [GF10] captured insights, a term which carries several definitions in the visualization community depending on the context [CZGR09]. In their study, Green et al. [GF10] define insights as "items or concepts learned or added to the user's knowledge base." They found that *external locus of control*, *introversion*, and low scores on the *neuroticism* scale mapped to more insights with their studied visualizations.

## 7.5. Participants

More than half (18 out of 29) of the studies recruited local college students (both undergraduates and graduate students), faculty or staff members as user study participants. 8 out of 29 studies recruited test subjects online, primarily from crowdsourcing platforms such as Amazon Mechanical Turk (a platform to recruit user study participants; see [PCI10] for more information about this service). The remaining studies did not report how they recruited participants.

Conducting user studies with crowdsourced participants is a relatively recent phenomenon, and it is becoming increasingly popular [ZOC*12b]. Some advantages of recruiting study participants via crowdsourcing include lower costs, fast access to a large pool of people who are willing to participate in user studies, and the ability to canvass diverse groups of users [HB10]. It might also reduce or avoid *experimenter bias* [PCI10, MDF12]. However, collecting data from crowdsourced experiments might raise concerns about data quality [OYC15, OCZC15, MHCV19]. Some participants may not take the experiments seriously and do something else while participating in a study [ZOC*12b]. Consequently, their responses could be poorly constructed, and hence invalid or valueless. For example, an uninterested participant might randomly pick answers during a session of multiple-choice questions. Some researchers also believe that it is possible for some participants to cheat by using search engines on their computers [ZOC*12b]. We observe that researchers commonly discard the data they deem problematic when they process experimental data from crowdsourced experiments [OCZC15, MHCV19]. Some researchers suggest that preferable conditions for crowdsourced experiments include studies with clear ground truth labels and incentives for speed and correctness [KZ10], and experimental tasks that are not severely affected by computing setups [ZOC*12b].

In addition, a study done by Kosara and Ziemkiewicz [KZ10] indicated that there were differences between student subjects and participants recruited from Amazon Mechanical Turk (MTurk). Compared to students, users recruited from MTurk had a significantly wider range of ages, a more balanced gender ratio, and different personalities (e.g., lower *agreeableness*, lower *extraversion* and higher *neuroticism*). Overall, it is advisable to keep these differences in mind when deciding whether an experiment should be conducted via crowdsourcing or not. In addition, more future research is needed to explore designs and techniques that improve the ecological validity of lab studies, because students belong to a narrow and selected subset of the general population.

## 8. Opportunities for Future Research

Visualization users differ greatly in experiences, backgrounds, personalities and cognitive abilities, yet visualizations, much like other software products, continue to be designed for a single ideal user. It would be clearly impractical to design each visualization for an individual user. However, knowledge of broad differences between user groups could be used to guide design for specific domains and to suggest multiple analysis modes or customization options in a single system. The body of work on individual differences in visualization provides a foundation for achieving this goal. However, successfully translating research to practice warrants more work. It is currently difficult for ordinary developers, with no background in visualization or social science research, to identify potential issues with their design choices. Perhaps the best-supported cognitive trait is color vision deficiency. There exist several designer tools for testing or verifying the color inclusiveness of a design [Cob, Cola] or for selecting palettes that are colorblind safe [Colb]. A key future direction is to enable practitioners, with no individual differences research background, to foresee the effect of their designs. More investigation is needed so that we can provide clear guidelines for research and practitioners, and success in the research agenda could transform how we evaluate and design visualizations for different user groups, tasks, and domains. There are many open questions and challenges.

### 8.1. Automatically Inferring Traits from Interaction

The research projects in this survey all used psychological surveys to estimate a person's cognitive traits. In real-world scenar-ios, however, it may be unrealistic to expect users to be subjected to a deluge of forms. Discovering new and unobtrusive methods to capture *cognitive state*, *cognitive trait*, and *experience/bias* will ultimately drive research in individual cognitive differences. For example, Brown et al. [BOZ*14] showed how we might detect user attributes by analyzing their click stream data, and others have demonstrated similar successes using eye tracking data [SCC13, CLRT17, TCC19]. In the broader visualization community, we have seen increased interest in developing algorithms to model users' behavior and in investigating how we can use these techniques to improve visualization tools (examples include [DC17], [OGW19] and [WBFE17]).

Although the research on user modeling and individual differences have largely been separate, analyzing their intersection could open the doors for many exciting future work. For instance, analyzing the portions of the *data* explored by the user can indicate a user's expertise and biases [WBFE17]. Brown et al. [BOZ*14] showed that analyzing *actions* (e.g., pans and zooms) uncovered differences that were mediated by user's *locus of control* scores and personality traits [BOZ*14]. Other work demonstrates that tracking *visual attention* via eye-gaze can reveal differences in people with varying *perceptual speed* and visual working memory [SCC13]. Therefore, to successfully infer individual traits, future work must consider a comprehensive set of encodings that include actions, data, and visual features. The ability to automatically infer personality traits and individual characteristics will open many opportunities for tailoring visualization systems to better suit the user. However, bridging the gap between visualization and personality psychology can raise serious privacy concerns. It is important to be aware of the potential ethical challenges ahead, and take socially-responsive steps to mitigate the effects.

### 8.2. Individual Differences and Open-Ended Tasks

Task design is critical to the success of an evaluation [Mun09], and researchers have created taxonomies for the types of tasks and interactions that are feasible for a given visualization (for example [AES05], [Shn96], [YaKSJ07], and [ZF98]). For future work, it is essential to recognize that "exploration" as a task carries several different meanings. Recent work by Battle and Heer [BH19] distinguishes between bottom-up exploration and top-down exploration. Bottom-up explorations "are driven in reaction to the data" [AZL*18] or "may be triggered by salient visual cues" [LH14]. This type of exploration is open-ended and the user's instincts largely drives the interactions. Top-down explorations, on the other hand, are based on high-level goals or hypotheses [BH19, GZ09, LH14].

One shortcoming of the prior work that investigates how individual traits impact exploration paths is they study only goal-driven, top-down exploration tasks. Because of this limitation, we know only the effect that individual traits have on interactions for top-down exploratory data analysis with short study duration. We need systematic studies to investigate the correlation between task types and patterns of interactions, and how individual traits may mediate observations over time. One possibility for expanding the body of literature is to investigate the impact of personality for more open-ended visual analytics tasks. The VAST Challenge [CGW14], for example, produces synthetic data annually and the challenges are designed to reflect real-world tasks under realistic conditions. Additionally, personal visualization [HTA*14] can incorporate data for use in a personal context. Future work can leverage these rich datasets to observe longitudinal top-down and bottom-up exploration processes and to uncover patterns in personality groups.

### 8.3. Generalization Across Visualization Designs

When we take a closer look at the previous results, much of the observed patterns of behavior can be explained by local versus global level precedence in processing information. For instance, when searching a tree visualization, participants with an *external locus of control* (*Externals*) were more likely to perform a depth-first search while participants with an *internal locus of control* (*Internals*) were more likely to perform a breadth-first search [OYC15]. A depth-first search strategy suggests a local precedence information processing while a breadth-first search indicates attention to global features and their relationships. As a result, Ottley et al. [OYC15] demonstrated that *Externals* were faster and more accurate with indented tree visualization. It is possible that the design encourages a local exploration. Similarly, when searching for Waldo, we found that *Externals* were more likely to explore at a lower zoom level, paying attention to local features, while *Internals* tended to only zoom in when they believed they had identified the target. This preference for attending to local versus global features suggests a pattern of behavior that may generalize across visualization designs. Future work is needed to investigate the relationship between individual traits and processing precedence across different designs.

### 8.4. Adaptation to Individual Needs

One important advantage of understanding individual users' *cognitive traits*, and *biases* as a cohesive structure is that this opens up the possibility of developing adaptive, mixed-initiative visualization systems [TC05]. Principles for similar mixed-initiative systems were proposed in the HCI community by Horvitz [Hor99]. As noted by Thomas and Cook in *Illuminating the Path* [TC05], an important direction in advancing visual analytics research is the development of an automated, computational system that can assist a user in performing analytical tasks. However, most visualization systems today are designed in a one-size-fits-all fashion without the ability to adapt to different users' analytical needs into the design.

Creating such mixed-initiative visualization systems is particularly difficult as visualization are often designed to support complex thought and decision-making. Still, there is some evidence that successful adaptive systems can significantly improve a user's ability in performing various tasks. For example, Gotz and Wen [GW09] proposed a behavior-driven visualization recommendation system that infers visual analytic tasks in real-time and suggests visualizations that might support the task better. Other work demonstrated how we can detect and adapt to mitigate exploration biases [GSC16, LDH*19, WBFE17]. It is clear that adaptive systems can offer new possibilities for visualization research and development [GW09], but additional work is necessary to understand *how* and *when* a system should adapt to a user's needs. Specifi-cally, studies are needed in order to carefully map features of the user unto the visual or interaction encodings of the system.

### 9. Conclusion

The community has made great strides in identifying characteristics that could impact performance on visualization systems. However, this work is still in its infancy, and uncovering the correlation between traits, visual design, and tasks is only the first step. What is clear from the existing body of research is that a mismatch between cognitive traits and visualization design can result in a gulf of evaluation or execution [ND86]. *The gulf of execution* describes the difference between the user's intentions and how well the design supports their goals. *The gulf of evaluation* is the difference between the system's state and the user's perceived state of the system. Vicente et al. [VHW87] acknowledges this challenge: "Although the assay and the isolation phases locate the locus of the individual differences in specific task components and certain user characteristics, they do not provide the designer with enough information to predict whether or not a given accommodation scheme will be successful."

Egan and Gomez [EG85] proposed a methodology for reducing the impact of individual differences for a given interface. They recommended a three-phased strategy: assaying, isolating, and accommodating individual differences. Assaying involves identifying key characteristics, and isolating requires understanding the interaction between the user characteristics and specific task components. Finally, the accommodation phase calls for changing or eliminating the problematic tasks.

The vast majority of the existing work are still in the assaying and isolating phase, and the community has yet to provide techniques for accommodating the broad variety of visualization users. We believe that this manuscript can serve as a central resource for practitioners and researchers to learn about the landscape of research on individual differences in visualization use and the implications for design and evaluation. Moreover, we hope this study will inspire future work that completes the understanding of individual differences and visualization, and serve as a catalyst for next-generation data visualization tools that better support individual users.

### Acknowledgments

We thank Jesse Huang'19 (Washington University in St. Louis) for his help with data collection, and Ananda Montoly '22 (Smith College) for her work validating the paper classification. This project was supported in part by: the Laboratory for Analytic Sciences at North Carolina State University, The Boeing Company under award 2018-BRT-PA-332, and the National Science Foundation under Grant No. 1755734.